# Test samples and infrastructure for accelerator magnet developments

Stoyan Stoynev[1]

*Abstract*— LHC still does not operate at full energy due to training features in magnets. Higher field practical superconductors (Nb$_3$Sn) show much worse training behavior than NbTi. Most magnet performance issues relate to quenching, and it is this phenomenon and its dependencies that need to be understood for us to succeed addressing them in magnets. For pragmatic and scientific reasons underlying mechanisms are hard to investigate in real magnets. Available data suggest that emulating local conditions at quench location may be enough to arrive at a good understanding of what drives observed behavior. "Local" conditions imply emulation by small samples of cables/"stacks" powered within a facility with controllable conditions, including external magnetic and force fields. This paper argues about the necessity of this approach, while reminding current understanding and directions of quench studies, and pointing out inadequacy of available techniques. Improved facilities and change of perspective are required for consistent and affordable development.

## I. Introduction

LHC is undoubtfully the most important instrument the HEP community has on its disposal and is widely considered an amazing technological achievement. At the core of this instrument lay accelerator magnets based on NbTi technology – a technology well developed and widely used in the commercial sector too. Yet, a decade after the accelerator started it still did not reach its design collision energy of 14 TeV [1] and it operated for years well below it. The reason for this is simple – magnet training [2], [3] (even though only a subset of the LHC magnets is mostly affected).

LHC upgrades and future colliders rely on new technologies allowing to reach higher operational fields and the most practical one now is based on Nb$_3$Sn conductor. Nb$_3$Sn superconducting properties were in fact discovered several years before NbTi. There are good reasons its development stalled - its mechanical properties, for one, are much worse than those of the very strong NbTi. As a result of this and mitigation measures, Nb$_3$Sn accelerator magnets of today, as a rule, show an unpleasant feature – much longer training than NbTi magnets.

After decades of magnet development, we are still unable to successfully deal with the problem of magnet training. The problem was known and acknowledged decades ago [4], there were and are continuous targeted attempts to solve it [5], [6], [7], [8], it is a top priority goal of the MDP [9]. Based on the many years of experience do we have the confidence it will be resolved soon enough, how? Moreover, a larger issue is quench behavior and performance limitations, and they too fall within the scope of the same question. In this paper, we discuss the phenomena involved and argue about steps the community should take to address them.

## II. Quench Training and More

### A. Training phenomenon, quenching

The manifestation of quench training is the semi-continuous rise in magnet current reached at consecutive power ramps before a natural quench breaks the ramping. A quench current plateau is eventually reached, eventually at design field. We talk about magnet quench training because this is what we observe directly. We see manifestations of phenomena, correlations between controlled quantities and behavior. Correlations are not enough to explain relations, we need to dig deeper.

It is probably reasonable to expect that processes during magnet powering and protection, especially large stress changes, after quenching affect the whole conductor – what-ever happens in one coil will be seen by the others. However, data do not support such a wide range statement. While variations of coil and magnet designs exist, most SC accelerator magnets are dipoles or quadrupoles, two or four "identical" coils per magnet, respectively. A special R&D object is a "mirror" magnet using only one coil with the rest replaced by iron ("mirror"). Although stress and field characteristics are not "mirrored" perfectly it is instructive to compare training behavior between coils. As shown in [10], with somewhat limited statistics, coils in "mirrors" and coils in "regular" configurations train similarly, i.e., "coil training" is better approximation than "magnet training". A focus on coils is also a simplification step to understand the underlying processes. If coils, which are typically double-layered, train independently - do layers train independently? Or are there sub-regions that train independently? How is "independently" even defined within a coil?

Sub-scale coils are often considered the smallest models for magnet development. However, multiple consecutive training quenches do happen in the same coil locations and can be even responsible for long stretches of the training curve if not defining it [11]. It is plausible and even logical to consider that the phenomenon could be driven by changes in local conditions

[1]S. Stoynev is with the Fermilab National Accelerator Laboratory, Batavia, IL, 60510, USA (e-mail: stoyan@fnal.gov).

alone – after all most natural quenches are believed to be caused by local energy release (a single minimum propagation zone developing). In that case a "sub-scale" model of those local conditions may be well suited to explore the phenomenon and pinpoint underlying reasons for it, possibly opening the way to mitigate adverse effects. Certainly, longer range current sharing effects could have non-trivial implications and could not be easily reproduced in any "sub-scale"; still they can be completely switched off and separately approximated in those experiments reducing the overall number of parameters to follow simultaneously.

While "training" is the use case we stick to, it is apparent that other quench dependencies and quenching itself can be subject to the same kind of scrutiny at "local" scale. Performance limitations, degradation, abnormal behaviors like reverse ramp rate quench current dependencies can all be studied at a smaller focused scale with conditions mimicking those in magnets. Defects could be introduced and studied while test samples will likely differ depending on research goals.

### B. Sub-scale models and training

Superconducting magnets are expensive, and the trend is going up. Almost all R&D objects are sub-scale magnets/prototypes, often of the order of one-two meters long, with otherwise the same structure as of a long magnet. Sometimes a coil(s) is (are) replaced by iron block(s). Or simplistic coils are built for specific studies (racetracks). All this works well for development but those are still expensive objects, we can build limited number of them, fabricating and testing them takes years and, yes, we still have problems figuring out the training. In fact, there is no firm understanding of what the exact causes are and what mitigations will help, beyond overall description of energy release sources and rich literature on possible solutions for magnets; there are concrete yet partial successes [12].

Because of cost considerations, magnet developments often rely on assumptions for magnet fabrication to move the field forward. There is still not much consensus on reliable testing, with direct link to magnets, on anything else but magnets. Recent CCTs development [13] reasonably proposed [14] CCTs to be a solution for training supported by a promising NbTi test [15]. Few years later the sub-scale magnets (~ 1 m long) in those experiments are replaced by even shorter CCT versions, themselves called subscale CCTs [16], intending, among other things, to resolve the extremely long training curves observed in Nb3Sn CCTs[12]. Activities are ongoing.

Hopefully the CCT subscales will resolve the familiar problem encountered, will move to the 1-2 m models to confirm the resolution and then to longer CCT coils. It is unclear which part of the potential solution will work for other accelerator magnet types – as said there is no consensus on a universal test model for "general" magnet R&D.

Since the Letter of Interest, which seeded the work on this paper, a new testing concept was proposed - BOX [17]. That work also contains an excellent set of references relevant here. The concept is a non-trivial development as it revolves around using an object, which is not a magnet, for magnet development. The basic concept is not new, per se: a cable "structure" for testing material and interfaces. The actual object though is a serpentine shaped cable, properly supported, immersed in external magnetic field, and it goes well beyond "cable testing" scope as currently understood. While authors emphasized the concept, they also showed initial results and discussed training curves of those devices. They also envisioned possible next steps: '…and to promote additional stress conditions by implementing in-situ compressive stresses on the broad face of the cable during powering…'.

Investigating the quench phenomenon, as well as many other aspects of magnet performance, in the context of accelerator magnets, naturally takes the shape of "sub-scale" development – ideally a "fast-turn-around", "low cost" solution with sizable statistical power to resolve built-in uncertainties. In search for "fast-turn-around" experiments sub-scale magnet models are often employed, indeed most of the R&D "short" magnet models are sub-scales of what a future design may look like. However, still smaller models are needed in practice. Training is a typical part of R&D magnet testing and the community have a good amount of data there; there are also some dedicated tests on smaller scale (including small solenoids) in self-field or in background field [5],[6]; there were tests on cables [18], there were tests conducted even on wires (for a subject review and references see [19]). While all those are steps in the right direction it can be questioned if such experiments, in particular on "fast-turn-around" models, can be directly related to training processes in actual coils. The logic would dictate to search for training patterns and relations in increasingly elaborate but logical structures – wire, wires, small magnets based on wire windings; cable, cables ("stack", with appropriate interfaces); small coils/magnets (various types); large coils/magnets (various types). As we care about the large (accelerator) magnets everything before them in the logical chain serves as an approximation and it needs to be determined at which point in the chain the approximation is good enough. The earlier we set the threshold the better (faster, cheaper). While experiments continue, the best we can rely on so far is still too expensive and slow to be considered adequate. This is hardly sustainable for R&D on quenching and training; and results will likely depend on magnet type used as a testing "model".

A better starting point is adopting a "local" reproduction of magnets, relatable to real situations in magnets; controlling and understanding of physics of quenches there gives a proper path to up-scaling of consistent performance characteristics and quench-group behavior. It is worth noting that the question of reliable up-scaling from small to large Nb$_3$Sn accelerator magnets is still not universally settled today as seen with on-going developments for LHC. The solution could not be to do more of the same, it never completely worked so far, and it is costly.

Except for a possible extension of BOX [17], there is no dedicated device that allows to test fully powered (Nb$_3$Sn) cables/"stacks" inside large magnetic field under pressure for the purpose of training exploration and quenching. There are devices for cable testing (like FRESCA2 [20]) though it is not clear if they can be used for dedicated quench or quench training studies. It is much better to design infrastructure that is more flexible, specifically conceived with training/quench studies in mind, that is, ideally, configurable for testing various samples. The byproduct of such a facility is that it is by construction also a cable testing one. To design a dedicated facility for cable/"stack" studies, pressure devices for wires (that is, a single strand is typically powered) inside magnetic field can serve as a model to upgrade [21] along with cable testing facilities themselves.

## III. Infrastructure for Magnet Development

### A. Emulation of local conditions

Conditions that govern or affect quenches in magnets are relatively limited in number. The main ones regard conductor (including design) and interfaces together with insulation material, current, magnetic field, pressure (force), temperature, cooling. All of them can be designed to be controllable in a small volume/area, within some uncertainty. Control is of utmost importance and variability of conditions, independently on each other, is a goal to pursue. It is to be noted that "locality" can also be a condition, in a sense that samples can vary in size and complexity. While the size of a minimum propagation zone at given environment is the natural scale for quench studies and "large enough" can be defined through that scale, properly designed longer/larger samples could test larger-scale effects on "locality".

The relative simplicity of such setups makes them perfectly suited for computer simulation reproduction. Simulation frameworks developed for magnets (like [22], [23], [24]), along with wide-use simulation tools, could be utilized to provide quite solid basis for such studies. Various models can be tested, software tuned and benchmarked. Simulations shall interactively yield to better designs of "local" samples and apparatuses, and eventually to better design and understanding of larger objects behavior.

### B. Test stands and samples

The BOX experiment [17], as the name itself suggests, had relatively narrow scope initially and its potential is likely to become more appreciated over time. Promising results are starting to appear [25]. In the context of this section, "BOX" is a "sample" which is utilized within an external structure (test stand) not designed with it in mind. With some modifications BOX could nearly give the practical example of a "sample" and a facility that can address the problems discussed so far. However, the fact remains that it fits within a facility not designed to provide flexibility (independent and comprehensive variation of key parameters) for testing. The general conception approaches the problem from the wrong end which is still better than not approaching it at all. To get maximum return a test stand designed with flexibility to test various samples at various conditions in mind is necessary. This does not mean it should be designed from scratch as "components" may be already available. It does mean comprehensive design studies are due with options evaluated. Those include magnet (type, field), mechanical press or presses (may also be based on interaction with another magnet), geometry of the system (magnet-press-sample), sample and extensions (limitations), cooling supply, multi-dimensional considerations (of parameters), degrees of control. The goal is to build a system that is able to reproduce "local" conductor conditions, as inferred, nearly anywhere in a real magnet and, better, different type of magnets.

The primary goal of such a facility is to enable studies on small superconducting samples (cables, "stacks") relatable to large magnets. It will allow to manipulate quenching and quench training in those samples at certain controllable conditions. Studies on the quench phenomenon and quench dependencies are much faster and easier on small samples, reliable simulations of those are to be expected. Large statistics and a limited sample area to study will contribute to solid results to be compared to real magnet behavior and features. Eventually, we will be able to reproduce patterns of both quench development and quench dependencies in large magnets on this small scale and reverse engineer coil designs to minimize unwanted effects. Simultaneously, the fast-turn-around models, the "stacks", are to be used for material, conductor, and interface studies in closer to real conditions environment. Over time this approach not only resolves problems but also saves valuable resources.

A main drawback of this approach may be considered the necessity to invest and concentrate efforts in non-accelerator-magnet development – mass fabrications and analysis of "stacks" and up-front investment of relevant facilities, as a start. If our community does not recognize the problems it has to resolve are not purely engineering but also scientific then we will continue hopping from technology to technology without ever reaching the true limits of any in practice. If we instead do recognize the true nature of issues, then a complex problem needs to be discerned into comprehensible pieces and resolved meticulously. Only then we can understand what makes magnet behaviors special.

## IV. Conclusion

Learning from the past suggests root problems in development of new technologies for accelerator magnets should be addressed at early stages. The community is past those stages in $Nb_3Sn$ R&D. A redirection of efforts is needed in order to avoid more costly disappointments in future. To understand persistent issues with magnet quenching it is necessary to investigate underlying mechanisms and build on this knowledge step by step. To do that with physical objects one needs statistics, reproducibility, precisely controllable environment, cost efficiency, comprehensive diagnostics, solid modeling support, fast turn-around times. Adopting configurable samples (cables/"stacks") that could be powered inside external magnetic and external force fields as part of dedicated test facilities could meet those needs. Such an approach should target emulation of "local" conditions in magnets. We have to learn how to control and influence behavior. So far results from complex structures (magnets) are not promising nor they point to a viable path forward. The proposed dedicated facilities and concepts do.